\documentclass[prl,twocolumn,showpacs,superscriptaddress,preprintnumbers,amssymb]{revtex4}
\usepackage{graphicx}
\usepackage{dcolumn}
\usepackage{bm}
\usepackage{wasysym}
\renewcommand{\v}[1]{{\bf #1}}
\newcommand{\sign}{{\rm sign}}
\newcommand{\M}{{\cal{M}}}
\def\vol{{\rm Vol}}
\newcommand{\s}{{\sigma}}
\newcommand{\psib}{{\bar{\psi}}}
\newcommand{\rb}{{\bar{\rho}}}
\newcommand{\phib}{{\bar{\phi}}}
\newcommand{\dt}{{\Delta}}
\newcommand{\w}{{\omega}}
\newcommand{\zh}{{\hat{z}}}
\newcommand{\qh}{{\hat{q}}}
\newcommand{\gr}{{\nabla}}
\newcommand{\beq}{\begin{equation}}
	\newcommand{\eeq}{\end{equation}}
\newcommand{\beqn}{\begin{eqnarray}}
	\newcommand{\eeqn}{\end{eqnarray}}
\newcommand{\nn}{\nonumber\\}
\newcommand{\Eq}[1]{Eq.~(\ref{#1})}
\newcommand{\p}{\partial}
\newcommand{\pt}{\partial_t}
\newcommand{\sgn}{{\rm sgn}}
\renewcommand{\t}[1]{{\tilde #1}}
\newcommand{\ua}{\uparrow}
\newcommand{\da}{\downarrow}
\newcommand{\ra}{\rightarrow}
\newcommand{\etc}{{\it etc~}}
\newcommand{\etal}{{\it etal~}}
\newcommand{\veps}{\varepsilon}
\newcommand{\vphi}{\varphi}
\newcommand{\cA}{ {\cal A} }
\newcommand{\cB}{ {\cal B} }
\newcommand{\cD}{ {\cal D} }
\newcommand{\cE}{ {\cal E} }
\newcommand{\cG}{ {\cal G} }
\newcommand{\cH}{ {\cal H} }
\newcommand{\cK}{ {\cal K} }
\newcommand{\cL}{ {\cal L} }
\newcommand{\cP}{ {\cal P} }
\newcommand{\cS}{ {\cal S} }
\newcommand{\cV}{ {\cal V} }
\newcommand{\ii}{\mathrm{i}}
\newcommand{\dd}{\mathrm{d}}
\newcommand{\bx}{\mathbf{x}}
\newcommand{\by}{\mathbf{y}}
\newcommand{\be}{\mathbf{e}}
\newcommand{\ve}{\vec{e}}
\newcommand{\he}{\hat{\vec{e}}}
\newcommand{\ba}{\mathbf{a}}
\newcommand{\va}{\vec{a}}
\newcommand{\ha}{\hat{\vec{a}}}
\newcommand{\bb}{\mathbf{b}}
\newcommand{\hb}{\hat{\vec{b}}}
\newcommand{\hn}{\hat{n}}
\newcommand{\SO}{\mathrm{SO}}
\renewcommand{\O}{\mathrm{O}}
\newcommand{\SU}{\mathrm{SU}}
\newcommand{\U}{\mathrm{U}}
\newcommand{\Sp}{\mathrm{Sp}}
\newcommand{\Gr}{\mathrm{Gr}}
\newcommand{\MF}{\mathrm{MF}}
\newcommand{\FM}{\mathrm{FM}}
\newcommand{\tr}{\mathrm{tr}}
\newcommand{\Hopf}{\mathrm{Hopf}}
\newcommand{\vect}[1]{{\bm{#1}}}
\newcommand{\refcite}[1]{Ref.\,\cite{#1}}
\newcommand{\eqnref}[1]{Eq.\,\eqref{#1}}
\newcommand{\figref}[1]{Fig.\,\ref{#1}}
\newcommand{\tabref}[1]{Tab.\,\ref{#1}}
\newcommand{\secref}[1]{Sec.\,\ref{#1}}
\newcommand{\appref}[1]{Appendix~\ref{#1}}
\newcommand{\nts}[1]{[\emph{\color{blue}{#1}}]}
\newcommand{\mat}[1]{\left(\begin{smallmatrix}#1\end{smallmatrix}\right)}
\renewcommand{\d}{\partial}
\newcommand{\nop}[2]{n^{#2}(#1,\bar{#1})}
\newcommand{\vop}[1]{v(#1,\bar{#1})}
\newcommand{\bbfamily}{\fontencoding{U}\fontfamily{bbold}\selectfont}
\DeclareMathAlphabet{\mathbbold}{U}{bbold}{m}{n}
\def\S{\textrm{S}}
\def\A{\textcolor{blue}{\mathcal{A}}}
\def\sgn{{\rm sgn}}
\def\tcrc{\textcolor{red}{citation}}
\def\Tr{\text{Tr}}
\def\Vol{\text{Vol}}
\def\SU{{\rm SU}}
\def\PSU{{\rm PSU}}
\def\U{{\rm U}}
\def\tT{\tilde{T}}
\def\ttT{\tilde{\tilde{T}}}
\def\dual{{\rm dual}}
\newcommand{\cmj}[1]{\textcolor{blue}{#1}}
\newcommand{\cx}[1]{\textcolor{red}{#1}}
\newcommand{\cC}{\mathcal{C}}
\newcommand{\lar}{\leftarrow}
\newcommand{\rar}{\rightarrow}
\newcommand{\lrar}{\leftrightarrow}

\begin{document}
	
	\title{Deconfined Quantum Critical Point with Non-locality}
	
	\author{Yichen Xu, Xiao-Chuan Wu, Cenke Xu}
	
	\affiliation{Department of Physics, University of California,
		Santa Barbara, CA 93106, USA}
	
	\begin{abstract}
		
		The deconfined quantum critical point (DQCP) between the N\'{e}el and valence bond solid (VBS) order was originally proposed in quantum spin systems with a local Hamiltonian. In the last few years analogues of DQCPs with nonlocal interactions have been explored, which can lead to rich possibilities. The nonlocal interactions can either arise from an instantaneous long range interaction in the Hamiltonian, or from gapless modes that reside in one higher spatial dimension. Here we consider another mechanism of generating nonlocal interactions by coupling the DQCP to the ``hot spots" of a Fermi surface. We demonstrate that at least within a substantial energy window, the physics of the DQCP is controlled by a new fixed point with dynamical exponent $z > 1$.
		
	\end{abstract}
	
	\maketitle
	
	{\it --- Introduction}
	
	A deconfined quantum critical point (DQCP) occurs between two phases that spontaneously break two different symmetries that do not contain each other as a subgroup. The original DQCP was proposed as a direct unfine-tuned continuous quantum phase transition between the collinear N\'{e}el and the valence bond solid (VBS) orders on the square lattice~\cite{deconfine1,deconfine2}. Various analogues of the original DQCP were studied, for example the transition between the superfluid and various density waves of a quantum boson system can be described in a similar framework as that of the DQCP with an easy-plane anisotropy~\cite{balentsboson,burkovbalents}; later the DQCP was also generalized to lattices where the spin order and VBS order both have different structure from the original DQCP~\cite{xutriangle}. In all these examples the two ordered phases separated by the DQCP break very different 0-form symmetries; but nowadays one can generalize the notion of DQCP to situations that involve higher-form symmetries. For example a direct transition between a magnetic order and a topological order can be viewed as a DQCP between a phase with spontaneous breaking of a 0-form symmetry and another phase with spontaneous breaking of a (emergent) 1-form symmetry~\cite{formsym0,formsym1,formsym2,formsym3,formsym4,formsym5,formsym6,formsym7,formsym8,mcgreevyreview}. In the past two decades, a lot of progress has been made towards understanding various aspects of the DQCP, including its connection to mixed 't Hooft anomaly and higher dimensional symmetry protected topological phases~\cite{senthilashvin}, as well as a duality web that connects different Lagrangian descriptions of the DQCP~\cite{lesikashvin,xudual,mrossdual,potterdual,seiberg2,deconfinedual,dualreview,mengdqcp2}, etc.
	
	Despite all the theoretical progresses, the nature of the original DQCP proposed on the square lattice has always remained controversial. Very encouraging evidences of DQCP were found in numerics on a $2D$ lattice quantum spin model dubbed the ``$J-Q$" model~\cite{JQ1,JQ2}, as well as loop models in the $3D$ Euclidean space~\cite{loopmodel1,loopmodel2}, but numerical simulations have also observed unusual scaling behaviors~\cite{JQ3,JQ4} and other complexities~\cite{mengdqcp}. Recently the DQCP has also been challenged by the ``conformal bootstrap" method of analyzing conformal field theories (CFT): the critical exponents obtained from numerical simulations seem incompatible with the bounds given by conformal bootstrap~\cite{bootstrap,bootstrap1}. Though these should not exclude the possibility that the DQCP still exists in other lattice models with critical exponents that are consistent with the conformal bootstrap bounds, a consensus on the nature of the DQCP awaits further efforts.
	
	In this work, rather than trying to address the infrared nature of the original DQCP, we explore a possible continuous quantum phase transition close to the originally proposed DQCP, starting with the transition between the easy-plane N\'{e}el order and the VBS order. In particular, we will discuss the effect of nonlocality on DQCP. Nonlocality of a system can directly arise from a long range instantaneous interaction in the Hamiltonian~\cite{sandvik1,sandvik2}, or from coupling to the gapless modes in one higher dimension, when the system is realized at the boundary of a bulk~\cite{groveredge,zhang1,zhang2,stefan1,stefan2,edgexu1,edgexu2,maxboundary,shang1,toldin1,toldin2,maboundary}. It was shown that, by coupling to the bulk quantum critical modes, the transition between the N\'{e}el and VBS order could be driven to a new fixed point~\cite{edgexu1,edgexu2}.
	
	Here we explore nonlocality arising from a more realistic mechanism. Nonlocality in space-time usually translates to nonanalyticity in the momentum-frequency space. It is well-known that, based on the Hertz-Millis theory~\cite{hertz,millis}, by coupling an order parameter $\phi$ to a Fermi surface, the dynamics of the order parameter acquires a singular contribution in the momentum-frequency space. In particular, when the order parameter carries a finite momentum that connects two ``hot spots" of the Fermi surface, after formally integrating out the fermions, the order parameter acquires a singular term $\sim  \sum_{\omega, \vect{q}} |\omega| |\phi_{\omega, \vect{q}}|^2 $. Within the framework of the Hertz-Millis theory, this singular term renders the original $\sum_{\omega, \vect{q}} \omega^2 |\phi_{\omega, \vect{q}}|^2 $ term in the Lagrangian irrelevant, and leads to a $z = 2$ Landau-Ginzburg theory of the order parameter $\phi$. But the effect of the coupling to the hot spots will be more complex in the case of DQCP, as the physical order parameter $\phi$ is now a composite operator of the deconfined degrees of freedom at the DQCP.  
	
	{\it --- Easy plane DQCP coupled with hot spots}
	
	Let us first inspect the easy-plane DQCP between the inplane N\'{e}el order and a VBS order on a square lattice. The order parameters involved in this transition include a two component inplane N\'{e}el order $(N_x, N_y)$ at momentum $(\pi,\pi)$, and a two component VBS order parameter $(V_x, V_y)$ at momentum $(\pi,0)$ and $(0,\pi)$ respectively. Since all these order parameters carry a finite momentum, in principle they would acquire a singular term in the form sketched above when the easy-plane DQCP occurs with a background Fermi surface, assuming their momenta connect hot spots of the Fermi surface. The Lagrangian that describes the easy-plane DQCP is an easy-plane CP$^1$ model, and it is known that this theory enjoys a self-duality~\cite{lesikashvin}, $i.e.$ the inplane N\'{e}el order parameter $(N_x, N_y)$ is a bilinear of the CP$^1$ field $(N_x, N_y) \sim (z^\dagger \sigma^x z, z^\dagger \sigma^y z)$, and the VBS order parameter along the $x$ and $y$ direction is a bilinear of the {\it dual} CP$^1$ field (the vortex of $z_1$ and $z_2$ respectively): $(V_x, V_y) \sim (v^\dagger \sigma^x v, v^\dagger \sigma^y v)$. Then after we integrate out the background Fermi surface according to the Hertz-Millis theory, the action that describes the transition becomes \beqn \mathcal{S} &=& \int d^2x d\tau \sum_{\alpha = 1, 2} |(\partial - \ii A) z_\alpha|^2 + r |z_\alpha|^2 + u |z_\alpha|^4 \cr\cr & + & \sum_{\omega, \vect{q}} \sum_{i = x, y} g |\omega| |(z^\dagger \sigma^i z)_{\omega, \vect{q}}|^2; \label{epcp1} \eeqn and the dual action reads \beqn \mathcal{S}_d &=& \int d^2x d\tau \sum_{\alpha = 1, 2} |(\partial - \ii \tilde{A}) v_\alpha|^2 + \tilde{r} |v_\alpha|^2 + \tilde{u} |v_\alpha|^4 \cr\cr & + & \sum_{\omega, \vect{q}} \sum_{i = x, y} \tilde{g} |\omega| |(v^\dagger \sigma^i v)_{\omega, \vect{q}}|^2, \label{epcp1d} \eeqn where $\tilde{r} = - r$. The actions above will be the starting point of our study; higher order singular terms beyond the Hertz-Millis theory that also arise from integrating out the background fermions will be briefly discussed later. In this work we will show that, although the bare values of $g$ and $\tilde{g}$ can differ, they actually flow to a fixed point where $g_\ast = \tilde{g}_\ast$. Hence this fixed point not only corresponds to a direct inplane N\'{e}el-to-VBS transition, this new fixed point still has the self-duality as the originally proposed easy-plane DQCP~\cite{lesikashvin}; but we do not make a statement about the presence of the enlarged emergent O(4) symmetry that can be perceived through the nonlinear Sigma model description of the easy-plane DQCP~\cite{senthilfisher}, as well as the duality web~\cite{xudual,mrossdual,potterdual,seiberg2,deconfinedual,dualreview,mengdqcp2}.
	
	In order to study the theory Eq.~\ref{epcp1} in a controllable fashion, we follow the standard procedure (see for example Ref.~\onlinecite{kaulsachdev,B2019}) by introducing the Hubbard-Stratonovich auxiliary fields $\lambda_\alpha$ and $\Phi^i$ to decompose the two quartic terms of $z_\alpha$, and consider the following large-$N$ generalization of Eq.~\ref{epcp1} at the critical point $r = 0$: \beqn \mathcal{S} &=& \int d^2x d\tau \sum_{a = 1}^{N} \sum_{\alpha = 1, 2} |(\partial - \ii A) z_{a, \alpha}|^2 + \ii\lambda_\alpha |z_{a, \alpha}|^2 \cr\cr &+& \ii \sum_{i = x, y} \Phi^i (z^\dagger_a \sigma^i z_a) \eeqn With large-$N$, the correlators of the Hubbard-Stratonovich fields, and the gauge field read \beqn && \langle \lambda_\alpha (\vec{q}) \lambda_{\alpha'}(- \vec{q}) \rangle = \frac{8}{N}|\vec{q}|\delta_{\alpha,\alpha'}, \cr\cr && \langle A_{\mu}(\vec{q}) A_{\nu}(-\vec{q}) \rangle = \frac{16}{2N}\left( \frac{\delta_{\mu\nu} - q_\mu q_\nu/q^2}{|q|}\right), \cr\cr && \langle \Phi^i(\vec{q}) \Phi^{i'}(- \vec{q}) \rangle = g |\nu| \delta_{i,i'}, \label{HS} \eeqn where $\vec{q} = (\nu, \vect{q})$. We assume that $g$ is at the order of $1/N$.
	
	\begin{figure}
		\includegraphics[width=6cm]{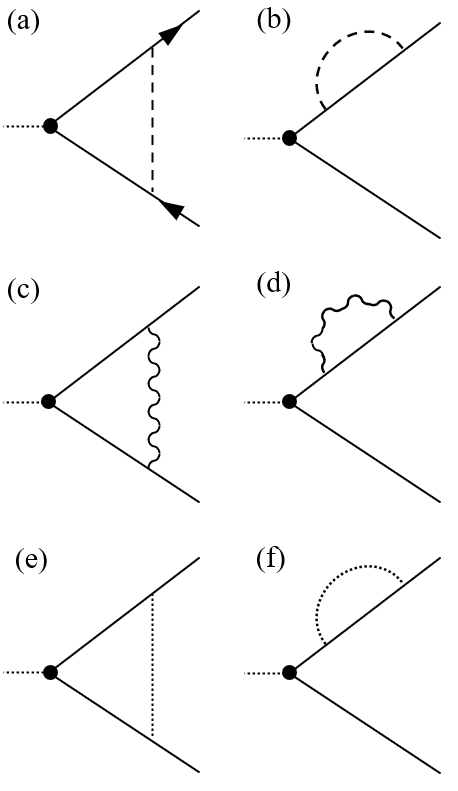}
		\caption{One-loop Feymann diagrams that contribute to $\beta(g)$. Here the solid, dashed, dotted and wavy lines represent the correlators of $z_{a,\alpha}$, $\lambda_\alpha$, $\Phi^i$ and gauge field $A_\mu$, respectively.}
		\label{diag1}
	\end{figure}
	
	
	We proceed by calculating the renormalization group (RG) flow of $g$ using the momentum-shell RG by integrating out the modes with momentum within $\Lambda/b < |k| < \Lambda$ (the calculations are repeated with the dimensional regularization as well); the most relevant Feynman diagrams are listed in \figref{diag1}. The diagrams (a-d) are the standard contributions to the leading order $1/N$ expansion of the CP$^{N-1}$ model~\cite{B2019}. The key of the calculation is the following: in the large-$N$ limit, the parameter $g$ is exactly marginal, as $z^\dagger_a \sigma^i z_a$ has scaling dimension $[z^\dagger_a \sigma^i z_a] =  1$ for $i = x,y$, while $\Phi^i$ has scaling dimension $[\Phi^i] = 2$. With finite $N$, the scaling dimension of $z^\dagger_a \sigma^i z_a$ receives a negative correction at the order of $1/N$: $[z^\dagger_a \sigma^i z_a] = 1 - 28/(3 \pi^2 N)$, which makes $g$ weakly relevant with large but finite $N$. Hence the beta function of $g$ should take the form \beqn \frac{dg}{d\ln b } = \beta(g) = \frac{56}{3\pi^2 N} g - C g^2. \eeqn When $C$ is positive and order unity, $g$ will flow to a fixed point at the order of $g_\ast \sim 1/N$.
	
	Diagram Fig.~\ref{diag1}(e) and (f) potentially contribute to the coefficient $C$ in the beta function above. Diagram (e) has vanishing contribution at the easy-plane DQCP under consideration right now, due to the matrix identity $\sum_{j=x,y}\sigma^j\sigma^i\sigma^j=0$ for $i=x,y$. Diagram (f) can be interpreted as the self-energy correction to $z_{a,\alpha}$ \beqn \Sigma(\omega, \vect{k})  &=& \sum_{j=x,y} g\sigma^j\sigma^j\int_{\Lambda/b}^\Lambda\frac{d^3 q}{(2\pi)^3}\frac{|\nu|}{(q+k)^2}\cr\cr &=& g\frac{\omega^2}{2\pi^2}\ln b\times \sigma^0+\dots. \eeqn Here $\vec{q} = (\nu, \vect{q})$, $\vec{k} = (\omega, \vect{k})$. The ellipses in the equation represent terms which do not contribute at order $\ln b$. This self-energy correction will modify the Gaussian part of the action of $z_{a,\alpha}$ to \beqn \mathcal{L} = z_{a,\alpha}^\ast \left(|\partial_\tau|^{2 \left(1 - g/(4\pi^2) \right)} - \partial_x^2 \right) z_{a,\alpha} + \cdots \eeqn This result implies that after coupling to the background Fermi surface, the space-time scaling of the original easy-plane DQCP is modified, which should now be
	\begin{equation}
		\tau \to b^{-z}\tau, \ \ \ \mathbf{x}\to b^{-1}\mathbf{x},
	\end{equation}
	where $z = 1 + \frac{g}{4\pi^2} + O(g^2)$ is the dynamical exponent. Here we remind the readers that, in the original Hertz-Millis theory, when an order parameter is coupled to hot spots of a Fermi surface, the Gaussian part of the Landau-Ginzburg theory of the order parameter has dynamical exponent $z = 2$. Here although the order parameter is a composite operator of $z_{a,\alpha}$, the dynamical scaling exponent $z$ is still modified due to its coupling to the Fermi surface.
	
	The wave function renormalization in diagram (f) is in fact equivalent to the modification of the space-time scaling, plus a correction to the scaling dimension of $z_{a,\alpha}$: $\Delta[z_{a,\alpha}] = g/(8\pi^2)$. Eventually the beta function of $g$ reads \beqn \beta(g)\equiv\frac{dg}{d\ln b} = \left(\frac{56}{3\pi^2 N}-\frac{g}{2\pi^2}\right)g, \eeqn where the first term arises from diagrams (a-d), while the second term is the additional wave-function renormalization from (f) as described above. Indeed, for $g>0$, the theory flows to a new fixed point $g_* = \frac{112}{3N} + O(\frac{1}{N^2})$. Several two-loop diagrams such as the Aslamazov-Larkin diagrams appear to be also at the $1/N$ order, but careful evaluation shows that these diagrams either do not contribute to the beta function as they do not lead to a logarithmic divergence, or their contributions cancel out with each other~\cite{B2019}.
	
	The same calculation applies to $\tilde{g}$ in Eq.~\ref{epcp1d}. Hence although the bare values of $g$ and $\tilde{g}$ in Eq.~\ref{epcp1} and Eq.~\ref{epcp1d} can be different, the RG equations above imply that they would flow to a fixed point where $g_\ast = \tilde{g}_\ast$. Hence at this fixed point the self-duality of the original easy-plane DQCP still holds. Another more technical note is that, the VBS order parameter $V_x \sim v^\dagger \sigma^x v$ is also the monopole operator of gauge field $A_\mu$ in Eq.~\ref{epcp1}, hence the $\tilde{g}$ term in Eq.~\ref{epcp1d} also corresponds to a correction to the action of the gauge field $A_\mu$. But since we expect $\tilde{g}$ to flow to a fixed point at order $1/N$, at the self-consistent level we can ignore this singular correction and use the gauge field propagator in the large-$N$ limit for our calculation.
	
	At this new RG fixed point, we obtain scaling dimensions for the following operators: \beqn &&[\lambda^+] = 2 - \frac{80}{3\pi^2N} - \frac{3g_*}{4\pi^2} = 2 - \frac{164}{3 \pi^2 N} + O(\frac{1}{N^2}),\cr\cr &&[\lambda^-] = 2 + \frac{16}{3\pi^2N} + \frac{g_*}{4\pi^2} = 2 + \frac{44}{\pi^2N} + O(\frac{1}{N^2}),\cr\cr &&[z^\dagger_a\sigma^{x,y}z_{a}]=1. \eeqn Here we have defined operators $\lambda^\pm=(\lambda_1\pm\lambda_2)/2$. Some two-loop diagrams like the ones considered in Ref.~\cite{wenwu} contribute to the evaluation of $[\lambda^+]$. The critical exponent $\nu$ is inferred from the scaling dimension of $\lambda^+$: \beqn \nu^{-1}=2 + z-[\lambda^+] = 1 + \frac{64}{\pi^2 N} + O(\frac{1}{N^2}). \eeqn These standard $1/N$ expansion may not be extremely reliable at the physically relevant case with small $N$, but the scaling dimensions of $z^\dagger_a\sigma^{x,y}z_{a}$ should be exactly $1$ at $g=g_*$, which is due to the fact that the beta function of $g$ evaluated at the fixed point $g_\ast$ vanishes: $\beta(g_*) = 0$. 
	
	We note that based on the Hertz-Millis theory there is another singular interaction $|\omega| |(z^\dagger \sigma^z z)_{\omega, \vect{q}}|^2$ that would also be generated by coupling the $z-$component of the N\'{e}el order to the background Fermi surface, but this term is irrelevant with the large-$N$ generalization of the easy-plane DQCP, as the scaling dimension of $z^\dagger \sigma^z z$ is 2 with large-$N$.
	
	We would also like to comment on the validity of the Hertz-Millis theory. It was noted in Ref.~\cite{hertzvalid} that, when we couple an order parameter $\phi$ to a Fermi surface, besides generating singular terms at the quadratic order of $\phi$, similar higher order terms $\sim \phi^n$ with space-time singularity is also generated after integrating out the fermions. It was shown in Ref.~\cite{hertzvalid} that direct power-counting suggests these higher order terms are marginal at the $z = 2$ Gaussian fixed point of the Hertz-Millis theory, hence it is no longer justified to ignore these terms. In fact, in our case, once we identify $\phi$ as $z^\dagger \sigma^i z$, the higher order terms pointed out in Ref.~\cite{hertzvalid} are still marginal at the new fixed point, since the scaling dimension $[z^\dagger \sigma^i z]$ is precisely 1. But this does not mean that the physics at the new fixed point we derived is not observable. Let us return to the original theory with a bosonic field $\phi$ coupled with $N_F$ copies of Fermi surfaces: \beqn
	\mathcal{L}_{BF} &=& \sum_{l=1}^{N_F} f^\dagger_{l1} (\partial_\tau - \ii\mathbf{v}_{1}\cdot\nabla)f_{l1} + f^\dagger_{l2} (\partial_\tau - \ii\mathbf{v}_{2}\cdot\nabla)f_{l2}\cr\cr
	& & + u\phi
	\left[\sum_l(f^\dagger_{l1} T f_{l2} + (1\lrar 2))\right],
	\eeqn where $1,2$ label two points of the Fermi surface connected by the momentum of $\phi$, and $T$ is a flavor matrix. The parameter $g$ in Eq.~\ref{epcp1} is about $g \sim N_F u^2$, and since $g \sim 1/N$, we need $u \sim \sqrt{1/(N N_F)}$. If we fix $N$, the higher order singular terms considered in Ref.~\cite{hertzvalid} will be at the order of $1/N_F^{n/2 - 1}$. Hence with large-$N_F$, although the ultimate fate of these higher order singular terms in the infrared limit is unclear, there could be a large energy window where the physics is controlled by the fixed point $g_\ast$ derived above. 
	
	We can also compute the self-energy of the fermions at the hot spot to the leading nontrivial order of $u$: \beqn \Sigma_{F}(\omega,\vect{k}) &\sim& 2u^2\sigma^0\int \frac{d\nu d^2\vect{q}}{(2\pi)^3} \  \frac{1}{\ii(\omega - \nu)-\mathbf{v}_2\cdot(\vect{k} - \vect{q})}\cr\cr &\times& \frac{1}{(\nu^2 + \vect{q}^2)^{1 - \eta/2}}. \eeqn 
	We have taken the correlator of the bosonic field $N_{x,y}(\vec{q}) \sim z^\dagger \sigma^{x,y} z$ to be $1/(\nu^2+ \vect{q}^2)^{1  - \eta/2}$, where $\eta$ is the anomalous dimension of the inplane N\'{e}el order parameter $N_{x,y}$ at the purely bosonic easy-plane DQCP. Carrying out the integral, we obtain \beqn \Sigma_F(\omega,0) \sim - \ii u^2 \sgn{(\omega)}|\omega|^{\eta}\sigma^0. \eeqn Generally we expect the fermions at the hot spots to have non-Fermi liquid like self-energy for a considerable energy window.

	
	{\it --- SU(2) invariant DQCP coupled with hot spots}
	
	Here we briefly discuss the SU(2)-invariant DQCP coupled to a background Fermi surface, which can be studied in the same way as the easy-plane case. Here we only need one Hubbard-Stratonovich field $\lambda^+$ to decompose the quartic term, and we obtain the following large-$N$ theory at the critical point:
	\beqn \mathcal{S} &=& \int d^2x d\tau \sum_{a = 1}^{N} \sum_{\alpha = 1, 2} |(\partial - \ii A) z_{a, \alpha}|^2 + \ii\lambda^+ |z_{a, \alpha}|^2 \cr\cr &+& \ii \sum_{i = x, y,z} \Phi^i (z^\dagger_a \sigma^i z_a). \eeqn Now the Fermi surfaces are coupled to all three components of the N\'eel order parameter $\vec{N}=z^\dagger\vec{\sigma}z$. The new dynamic exponent is now $z=1 + \frac{3g}{8\pi^2} + O(g^2)$, and the beta function of $g$ is
	\begin{equation}
		\beta(g)\equiv\frac{dg}{d\ln b} = \frac{16}{\pi^2 N}g - \frac{g^2}{4\pi^2}.
	\end{equation}
	If $g>0$, the theory flows to a new RG fixed point at $g_\ast = \frac{64}{N}$. At the new RG fixed point, we have the following scaling dimensions \beqn &&[\lambda^+] = 2 - \frac{24}{\pi^2N} - \frac{9g_*}{8\pi^2}=2 - \frac{96}{\pi^2N} + O(\frac{1}{N^2}),\cr\cr &&[z^\dagger_a\sigma^{x,y,z}z_{a}] = 1, \eeqn and the critical exponent \beqn \nu^{-1}=2 + z - [\lambda^+]=1 + \frac{120}{\pi^2 N} + O(\frac{1}{N^2}).
	\eeqn Again, the N\'eel order parameter has scaling dimension $[\vec{N}] = 1$ exactly at the new fixed point. Though it is not so convenient to directly compute the RG flow of the singular term of the VBS order parameter due to the lack of a dual Lagrangian for the SU(2) invariant DQCP, we expect the scaling dimension of the VBS order parameter should also be 1 at the new fixed point, which is the same as the N\'eel order parameter, and it is compatible with an emergent SO(5) symmetry as was suggested in previous literature for DQCP~\cite{senthilfisher,SO5,loopmodel1}.
	
	{\it --- Discussion}
	
	In this work we discussed the fate of the DQCP when it occurs with a background Fermi surface. We demonstrated that with a large number of copies of Fermi surfaces, there is a substantial energy window where the easy-plane DQCP is controlled by a self-dual fixed point with dynamical exponent $z > 1$. We did not pursue a full renormalization group analysis of the boson-fermion coupled theory, but such analysis like the ones discussed in Ref.~\cite{metlitskisdw,leesdw} when the order parameter $\phi$ is a composite operator of deconfined degrees of freedom is very much worth studying in the future. 
	
	Many insights of the DQCP, including the emergent symmetry, 't Hooft anomaly, as well as possible phase diagram and RG flow, can be gained from the nonlinear sigma model (NLSM) approach that unifies all the order parameters in one action~\cite{senthilfisher,xuludwig,xunlsm,wangnlsm,nahumlsm,assaadnlsm,stiefel}. The very key term in the NLSM is a topological term. In the future it is also worth to explore the consequence of coupling the DQCP to a Fermi surface using the NLSM formalism. 
	
	Besides the DQCP, our study is also meaningful to the interaction-driven Metal-insulator transition (MIT) where the insulator phase has certain density wave order. The basic formalism of the theory describing this MIT involves introducing bosonic partons that carry the electric charge, and fermionic partons that carry the spin. This MIT is interpreted as a superfluid-to-density wave transition of the charged bosonic parton sector~\cite{fractionalMIT,wignerMIT} (The ``superfluid" phase of the bosonic sector of the phase diagram corresponds to the metallic phase~\cite{senthilMIT}), which is also described by a CP$^{N-1}$ model in which the bosonic matter fields are vortices of the charged bosonic parton. There are multiple components of the vortex fields whose condensate corresponds to the degenerate density wave patterns of the insulator phase. When the density wave order parameter couples to the hot spots of the Fermi surface of the fermionic spinon sector, the same singular terms like the one considered in our current work will arise. Our study indicates that the physics at this MIT could be controlled by a new fixed point with dynamical exponent $z > 1$. 
	
	This work is supported by NSF Grant No. DMR-1920434, and the Simons Investigator program.
	
	\bibliography{NLDQCP}

\begin{thebibliography}{66}
\expandafter\ifx\csname natexlab\endcsname\relax\def\natexlab#1{#1}\fi
\expandafter\ifx\csname bibnamefont\endcsname\relax
  \def\bibnamefont#1{#1}\fi
\expandafter\ifx\csname bibfnamefont\endcsname\relax
  \def\bibfnamefont#1{#1}\fi
\expandafter\ifx\csname citenamefont\endcsname\relax
  \def\citenamefont#1{#1}\fi
\expandafter\ifx\csname url\endcsname\relax
  \def\url#1{\texttt{#1}}\fi
\expandafter\ifx\csname urlprefix\endcsname\relax\def\urlprefix{URL }\fi
\providecommand{\bibinfo}[2]{#2}
\providecommand{\eprint}[2][]{\url{#2}}

\bibitem[{\citenamefont{Senthil
  et~al.}(2004{\natexlab{a}})\citenamefont{Senthil, Vishwanath, Balents,
  Sachdev, and Fisher}}]{deconfine1}
\bibinfo{author}{\bibfnamefont{T.}~\bibnamefont{Senthil}},
  \bibinfo{author}{\bibfnamefont{A.}~\bibnamefont{Vishwanath}},
  \bibinfo{author}{\bibfnamefont{L.}~\bibnamefont{Balents}},
  \bibinfo{author}{\bibfnamefont{S.}~\bibnamefont{Sachdev}}, \bibnamefont{and}
  \bibinfo{author}{\bibfnamefont{M.~P.~A.} \bibnamefont{Fisher}},
  \bibinfo{journal}{Science} \textbf{\bibinfo{volume}{303}},
  \bibinfo{pages}{1490} (\bibinfo{year}{2004}{\natexlab{a}}).

\bibitem[{\citenamefont{Senthil
  et~al.}(2004{\natexlab{b}})\citenamefont{Senthil, Balents, Sachdev,
  Vishwanath, and Fisher}}]{deconfine2}
\bibinfo{author}{\bibfnamefont{T.}~\bibnamefont{Senthil}},
  \bibinfo{author}{\bibfnamefont{L.}~\bibnamefont{Balents}},
  \bibinfo{author}{\bibfnamefont{S.}~\bibnamefont{Sachdev}},
  \bibinfo{author}{\bibfnamefont{A.}~\bibnamefont{Vishwanath}},
  \bibnamefont{and} \bibinfo{author}{\bibfnamefont{M.~P.~A.}
  \bibnamefont{Fisher}}, \bibinfo{journal}{Phys. Rev. B}
  \textbf{\bibinfo{volume}{70}}, \bibinfo{pages}{144407}
  (\bibinfo{year}{2004}{\natexlab{b}}),
  \urlprefix\url{https://link.aps.org/doi/10.1103/PhysRevB.70.144407}.

\bibitem[{\citenamefont{Balents et~al.}(2005)\citenamefont{Balents, Bartosch,
  Burkov, Sachdev, and Sengupta}}]{balentsboson}
\bibinfo{author}{\bibfnamefont{L.}~\bibnamefont{Balents}},
  \bibinfo{author}{\bibfnamefont{L.}~\bibnamefont{Bartosch}},
  \bibinfo{author}{\bibfnamefont{A.}~\bibnamefont{Burkov}},
  \bibinfo{author}{\bibfnamefont{S.}~\bibnamefont{Sachdev}}, \bibnamefont{and}
  \bibinfo{author}{\bibfnamefont{K.}~\bibnamefont{Sengupta}},
  \bibinfo{journal}{Progress of Theoretical Physics Supplement}
  \textbf{\bibinfo{volume}{160}}, \bibinfo{pages}{314} (\bibinfo{year}{2005}),
  ISSN \bibinfo{issn}{0375-9687},
  \eprint{https://academic.oup.com/ptps/article-pdf/doi/10.1143/PTPS.160.314/5164035/160-314.pdf},
  \urlprefix\url{https://doi.org/10.1143/PTPS.160.314}.

\bibitem[{\citenamefont{Burkov and Balents}(2005)}]{burkovbalents}
\bibinfo{author}{\bibfnamefont{A.~A.} \bibnamefont{Burkov}} \bibnamefont{and}
  \bibinfo{author}{\bibfnamefont{L.}~\bibnamefont{Balents}},
  \bibinfo{journal}{Phys. Rev. B} \textbf{\bibinfo{volume}{72}},
  \bibinfo{pages}{134502} (\bibinfo{year}{2005}),
  \urlprefix\url{https://link.aps.org/doi/10.1103/PhysRevB.72.134502}.

\bibitem[{\citenamefont{Jian et~al.}(2018)\citenamefont{Jian, Thomson,
  Rasmussen, Bi, and Xu}}]{xutriangle}
\bibinfo{author}{\bibfnamefont{C.-M.} \bibnamefont{Jian}},
  \bibinfo{author}{\bibfnamefont{A.}~\bibnamefont{Thomson}},
  \bibinfo{author}{\bibfnamefont{A.}~\bibnamefont{Rasmussen}},
  \bibinfo{author}{\bibfnamefont{Z.}~\bibnamefont{Bi}}, \bibnamefont{and}
  \bibinfo{author}{\bibfnamefont{C.}~\bibnamefont{Xu}}, \bibinfo{journal}{Phys.
  Rev. B} \textbf{\bibinfo{volume}{97}}, \bibinfo{pages}{195115}
  (\bibinfo{year}{2018}),
  \urlprefix\url{https://link.aps.org/doi/10.1103/PhysRevB.97.195115}.

\bibitem[{\citenamefont{Nussinov and Ortiz}(2009)}]{formsym0}
\bibinfo{author}{\bibfnamefont{Z.}~\bibnamefont{Nussinov}} \bibnamefont{and}
  \bibinfo{author}{\bibfnamefont{G.}~\bibnamefont{Ortiz}},
  \bibinfo{journal}{Annals of Physics} \textbf{\bibinfo{volume}{324}},
  \bibinfo{pages}{977} (\bibinfo{year}{2009}), ISSN \bibinfo{issn}{0003-4916},
  \urlprefix\url{http://dx.doi.org/10.1016/j.aop.2008.11.002}.

\bibitem[{\citenamefont{Aharony et~al.}(2013)\citenamefont{Aharony, Seiberg,
  and Tachikawa}}]{formsym1}
\bibinfo{author}{\bibfnamefont{O.}~\bibnamefont{Aharony}},
  \bibinfo{author}{\bibfnamefont{N.}~\bibnamefont{Seiberg}}, \bibnamefont{and}
  \bibinfo{author}{\bibfnamefont{Y.}~\bibnamefont{Tachikawa}},
  \bibinfo{journal}{Journal of High Energy Physics}
  \textbf{\bibinfo{volume}{2013}} (\bibinfo{year}{2013}), ISSN
  \bibinfo{issn}{1029-8479},
  \urlprefix\url{http://dx.doi.org/10.1007/JHEP08(2013)115}.

\bibitem[{\citenamefont{Gukov and Kapustin}(2013)}]{formsym2}
\bibinfo{author}{\bibfnamefont{S.}~\bibnamefont{Gukov}} \bibnamefont{and}
  \bibinfo{author}{\bibfnamefont{A.}~\bibnamefont{Kapustin}}
  (\bibinfo{year}{2013}), \eprint{1307.4793}.

\bibitem[{\citenamefont{Kapustin and Thorngren}(2013{\natexlab{a}})}]{formsym3}
\bibinfo{author}{\bibfnamefont{A.}~\bibnamefont{Kapustin}} \bibnamefont{and}
  \bibinfo{author}{\bibfnamefont{R.}~\bibnamefont{Thorngren}}
  (\bibinfo{year}{2013}{\natexlab{a}}), \eprint{1308.2926}.

\bibitem[{\citenamefont{Kapustin and Thorngren}(2013{\natexlab{b}})}]{formsym4}
\bibinfo{author}{\bibfnamefont{A.}~\bibnamefont{Kapustin}} \bibnamefont{and}
  \bibinfo{author}{\bibfnamefont{R.}~\bibnamefont{Thorngren}}
  (\bibinfo{year}{2013}{\natexlab{b}}), \eprint{1309.4721}.

\bibitem[{\citenamefont{Kapustin and Seiberg}(2014)}]{formsym5}
\bibinfo{author}{\bibfnamefont{A.}~\bibnamefont{Kapustin}} \bibnamefont{and}
  \bibinfo{author}{\bibfnamefont{N.}~\bibnamefont{Seiberg}},
  \bibinfo{journal}{Journal of High Energy Physics}
  \textbf{\bibinfo{volume}{2014}} (\bibinfo{year}{2014}), ISSN
  \bibinfo{issn}{1029-8479},
  \urlprefix\url{http://dx.doi.org/10.1007/JHEP04(2014)001}.

\bibitem[{\citenamefont{Gaiotto et~al.}(2015)\citenamefont{Gaiotto, Kapustin,
  Seiberg, and Willett}}]{formsym6}
\bibinfo{author}{\bibfnamefont{D.}~\bibnamefont{Gaiotto}},
  \bibinfo{author}{\bibfnamefont{A.}~\bibnamefont{Kapustin}},
  \bibinfo{author}{\bibfnamefont{N.}~\bibnamefont{Seiberg}}, \bibnamefont{and}
  \bibinfo{author}{\bibfnamefont{B.}~\bibnamefont{Willett}},
  \bibinfo{journal}{Journal of High Energy Physics}
  \textbf{\bibinfo{volume}{2015}} (\bibinfo{year}{2015}), ISSN
  \bibinfo{issn}{1029-8479},
  \urlprefix\url{http://dx.doi.org/10.1007/JHEP02(2015)172}.

\bibitem[{\citenamefont{Hsin et~al.}(2019)\citenamefont{Hsin, Lam, and
  Seiberg}}]{formsym7}
\bibinfo{author}{\bibfnamefont{P.-S.} \bibnamefont{Hsin}},
  \bibinfo{author}{\bibfnamefont{H.~T.} \bibnamefont{Lam}}, \bibnamefont{and}
  \bibinfo{author}{\bibfnamefont{N.}~\bibnamefont{Seiberg}},
  \bibinfo{journal}{SciPost Physics} \textbf{\bibinfo{volume}{6}}
  (\bibinfo{year}{2019}), ISSN \bibinfo{issn}{2542-4653},
  \urlprefix\url{http://dx.doi.org/10.21468/SciPostPhys.6.3.039}.

\bibitem[{\citenamefont{Seiberg}(2020)}]{formsym8}
\bibinfo{author}{\bibfnamefont{N.}~\bibnamefont{Seiberg}},
  \bibinfo{journal}{SciPost Physics} \textbf{\bibinfo{volume}{8}}
  (\bibinfo{year}{2020}), ISSN \bibinfo{issn}{2542-4653},
  \urlprefix\url{http://dx.doi.org/10.21468/SciPostPhys.8.4.050}.

\bibitem[{\citenamefont{McGreevy}(2022)}]{mcgreevyreview}
\bibinfo{author}{\bibfnamefont{J.}~\bibnamefont{McGreevy}},
  \emph{\bibinfo{title}{Generalized symmetries in condensed matter}}
  (\bibinfo{year}{2022}), \urlprefix\url{https://arxiv.org/abs/2204.03045}.

\bibitem[{\citenamefont{Vishwanath and Senthil}(2013)}]{senthilashvin}
\bibinfo{author}{\bibfnamefont{A.}~\bibnamefont{Vishwanath}} \bibnamefont{and}
  \bibinfo{author}{\bibfnamefont{T.}~\bibnamefont{Senthil}},
  \bibinfo{journal}{Phys. Rev. X} \textbf{\bibinfo{volume}{3}},
  \bibinfo{pages}{011016} (\bibinfo{year}{2013}),
  \urlprefix\url{https://link.aps.org/doi/10.1103/PhysRevX.3.011016}.

\bibitem[{\citenamefont{Motrunich and Vishwanath}(2004)}]{lesikashvin}
\bibinfo{author}{\bibfnamefont{O.~I.} \bibnamefont{Motrunich}}
  \bibnamefont{and}
  \bibinfo{author}{\bibfnamefont{A.}~\bibnamefont{Vishwanath}},
  \bibinfo{journal}{Phys. Rev. B} \textbf{\bibinfo{volume}{70}},
  \bibinfo{pages}{075104} (\bibinfo{year}{2004}),
  \urlprefix\url{https://link.aps.org/doi/10.1103/PhysRevB.70.075104}.

\bibitem[{\citenamefont{Xu and You}(2015)}]{xudual}
\bibinfo{author}{\bibfnamefont{C.}~\bibnamefont{Xu}} \bibnamefont{and}
  \bibinfo{author}{\bibfnamefont{Y.-Z.} \bibnamefont{You}},
  \bibinfo{journal}{Phys. Rev. B} \textbf{\bibinfo{volume}{92}},
  \bibinfo{pages}{220416} (\bibinfo{year}{2015}),
  \urlprefix\url{https://link.aps.org/doi/10.1103/PhysRevB.92.220416}.

\bibitem[{\citenamefont{Mross et~al.}(2016)\citenamefont{Mross, Alicea, and
  Motrunich}}]{mrossdual}
\bibinfo{author}{\bibfnamefont{D.~F.} \bibnamefont{Mross}},
  \bibinfo{author}{\bibfnamefont{J.}~\bibnamefont{Alicea}}, \bibnamefont{and}
  \bibinfo{author}{\bibfnamefont{O.~I.} \bibnamefont{Motrunich}},
  \bibinfo{journal}{Phys. Rev. Lett.} \textbf{\bibinfo{volume}{117}},
  \bibinfo{pages}{016802} (\bibinfo{year}{2016}),
  \urlprefix\url{https://link.aps.org/doi/10.1103/PhysRevLett.117.016802}.

\bibitem[{\citenamefont{Potter et~al.}(2017)\citenamefont{Potter, Wang,
  Metlitski, and Vishwanath}}]{potterdual}
\bibinfo{author}{\bibfnamefont{A.~C.} \bibnamefont{Potter}},
  \bibinfo{author}{\bibfnamefont{C.}~\bibnamefont{Wang}},
  \bibinfo{author}{\bibfnamefont{M.~A.} \bibnamefont{Metlitski}},
  \bibnamefont{and}
  \bibinfo{author}{\bibfnamefont{A.}~\bibnamefont{Vishwanath}},
  \bibinfo{journal}{Phys. Rev. B} \textbf{\bibinfo{volume}{96}},
  \bibinfo{pages}{235114} (\bibinfo{year}{2017}),
  \urlprefix\url{https://link.aps.org/doi/10.1103/PhysRevB.96.235114}.

\bibitem[{\citenamefont{Hsin and Seiberg}(2016)}]{seiberg2}
\bibinfo{author}{\bibfnamefont{P.-S.} \bibnamefont{Hsin}} \bibnamefont{and}
  \bibinfo{author}{\bibfnamefont{N.}~\bibnamefont{Seiberg}},
  \bibinfo{journal}{Journal of High Energy Physics}
  \textbf{\bibinfo{volume}{2016}}, \bibinfo{pages}{95} (\bibinfo{year}{2016}),
  ISSN \bibinfo{issn}{1029-8479},
  \urlprefix\url{http://dx.doi.org/10.1007/JHEP09(2016)095}.

\bibitem[{\citenamefont{Wang et~al.}(2017{\natexlab{a}})\citenamefont{Wang,
  Nahum, Metlitski, Xu, and Senthil}}]{deconfinedual}
\bibinfo{author}{\bibfnamefont{C.}~\bibnamefont{Wang}},
  \bibinfo{author}{\bibfnamefont{A.}~\bibnamefont{Nahum}},
  \bibinfo{author}{\bibfnamefont{M.~A.} \bibnamefont{Metlitski}},
  \bibinfo{author}{\bibfnamefont{C.}~\bibnamefont{Xu}}, \bibnamefont{and}
  \bibinfo{author}{\bibfnamefont{T.}~\bibnamefont{Senthil}},
  \bibinfo{journal}{Phys. Rev. X} \textbf{\bibinfo{volume}{7}},
  \bibinfo{pages}{031051} (\bibinfo{year}{2017}{\natexlab{a}}),
  \urlprefix\url{https://link.aps.org/doi/10.1103/PhysRevX.7.031051}.

\bibitem[{\citenamefont{Senthil et~al.}(2019)\citenamefont{Senthil, Son, Wang,
  and Xu}}]{dualreview}
\bibinfo{author}{\bibfnamefont{T.}~\bibnamefont{Senthil}},
  \bibinfo{author}{\bibfnamefont{D.~T.} \bibnamefont{Son}},
  \bibinfo{author}{\bibfnamefont{C.}~\bibnamefont{Wang}}, \bibnamefont{and}
  \bibinfo{author}{\bibfnamefont{C.}~\bibnamefont{Xu}},
  \bibinfo{journal}{Physics Reports} \textbf{\bibinfo{volume}{827}},
  \bibinfo{pages}{1 } (\bibinfo{year}{2019}), ISSN \bibinfo{issn}{0370-1573},
  \bibinfo{note}{duality between (2+1)d quantum critical points},
  \urlprefix\url{http://www.sciencedirect.com/science/article/pii/S0370157319302637}.

\bibitem[{\citenamefont{Qin et~al.}(2017)\citenamefont{Qin, He, You, Lu, Sen,
  Sandvik, Xu, and Meng}}]{mengdqcp2}
\bibinfo{author}{\bibfnamefont{Y.~Q.} \bibnamefont{Qin}},
  \bibinfo{author}{\bibfnamefont{Y.-Y.} \bibnamefont{He}},
  \bibinfo{author}{\bibfnamefont{Y.-Z.} \bibnamefont{You}},
  \bibinfo{author}{\bibfnamefont{Z.-Y.} \bibnamefont{Lu}},
  \bibinfo{author}{\bibfnamefont{A.}~\bibnamefont{Sen}},
  \bibinfo{author}{\bibfnamefont{A.~W.} \bibnamefont{Sandvik}},
  \bibinfo{author}{\bibfnamefont{C.}~\bibnamefont{Xu}}, \bibnamefont{and}
  \bibinfo{author}{\bibfnamefont{Z.~Y.} \bibnamefont{Meng}},
  \bibinfo{journal}{Phys. Rev. X} \textbf{\bibinfo{volume}{7}},
  \bibinfo{pages}{031052} (\bibinfo{year}{2017}),
  \urlprefix\url{https://link.aps.org/doi/10.1103/PhysRevX.7.031052}.

\bibitem[{\citenamefont{Sandvik}(2007)}]{JQ1}
\bibinfo{author}{\bibfnamefont{A.~W.} \bibnamefont{Sandvik}},
  \bibinfo{journal}{Phys. Rev. Lett.} \textbf{\bibinfo{volume}{98}},
  \bibinfo{pages}{227202} (\bibinfo{year}{2007}),
  \urlprefix\url{https://link.aps.org/doi/10.1103/PhysRevLett.98.227202}.

\bibitem[{\citenamefont{Melko and Kaul}(2008)}]{JQ2}
\bibinfo{author}{\bibfnamefont{R.~G.} \bibnamefont{Melko}} \bibnamefont{and}
  \bibinfo{author}{\bibfnamefont{R.~K.} \bibnamefont{Kaul}},
  \bibinfo{journal}{Phys. Rev. Lett.} \textbf{\bibinfo{volume}{100}},
  \bibinfo{pages}{017203} (\bibinfo{year}{2008}),
  \urlprefix\url{https://link.aps.org/doi/10.1103/PhysRevLett.100.017203}.

\bibitem[{\citenamefont{Nahum et~al.}(2015{\natexlab{a}})\citenamefont{Nahum,
  Serna, Chalker, Ortu\~no, and Somoza}}]{loopmodel1}
\bibinfo{author}{\bibfnamefont{A.}~\bibnamefont{Nahum}},
  \bibinfo{author}{\bibfnamefont{P.}~\bibnamefont{Serna}},
  \bibinfo{author}{\bibfnamefont{J.~T.} \bibnamefont{Chalker}},
  \bibinfo{author}{\bibfnamefont{M.}~\bibnamefont{Ortu\~no}}, \bibnamefont{and}
  \bibinfo{author}{\bibfnamefont{A.~M.} \bibnamefont{Somoza}},
  \bibinfo{journal}{Phys. Rev. Lett.} \textbf{\bibinfo{volume}{115}},
  \bibinfo{pages}{267203} (\bibinfo{year}{2015}{\natexlab{a}}),
  \urlprefix\url{https://link.aps.org/doi/10.1103/PhysRevLett.115.267203}.

\bibitem[{\citenamefont{Nahum et~al.}(2015{\natexlab{b}})\citenamefont{Nahum,
  Chalker, Serna, Ortu\~no, and Somoza}}]{loopmodel2}
\bibinfo{author}{\bibfnamefont{A.}~\bibnamefont{Nahum}},
  \bibinfo{author}{\bibfnamefont{J.~T.} \bibnamefont{Chalker}},
  \bibinfo{author}{\bibfnamefont{P.}~\bibnamefont{Serna}},
  \bibinfo{author}{\bibfnamefont{M.}~\bibnamefont{Ortu\~no}}, \bibnamefont{and}
  \bibinfo{author}{\bibfnamefont{A.~M.} \bibnamefont{Somoza}},
  \bibinfo{journal}{Phys. Rev. X} \textbf{\bibinfo{volume}{5}},
  \bibinfo{pages}{041048} (\bibinfo{year}{2015}{\natexlab{b}}),
  \urlprefix\url{https://link.aps.org/doi/10.1103/PhysRevX.5.041048}.

\bibitem[{\citenamefont{Sandvik}(2010{\natexlab{a}})}]{JQ3}
\bibinfo{author}{\bibfnamefont{A.~W.} \bibnamefont{Sandvik}},
  \bibinfo{journal}{Phys. Rev. Lett.} \textbf{\bibinfo{volume}{104}},
  \bibinfo{pages}{177201} (\bibinfo{year}{2010}{\natexlab{a}}),
  \urlprefix\url{https://link.aps.org/doi/10.1103/PhysRevLett.104.177201}.

\bibitem[{\citenamefont{Shao et~al.}(2016)\citenamefont{Shao, Guo, and
  Sandvik}}]{JQ4}
\bibinfo{author}{\bibfnamefont{H.}~\bibnamefont{Shao}},
  \bibinfo{author}{\bibfnamefont{W.}~\bibnamefont{Guo}}, \bibnamefont{and}
  \bibinfo{author}{\bibfnamefont{A.~W.} \bibnamefont{Sandvik}},
  \bibinfo{journal}{Science} \textbf{\bibinfo{volume}{352}},
  \bibinfo{pages}{213} (\bibinfo{year}{2016}),
  \urlprefix\url{https://doi.org/10.1126%2Fscience.aad5007}.

\bibitem[{\citenamefont{Zhao et~al.}(2022)\citenamefont{Zhao, Wang, Yan, Cheng,
  and Meng}}]{mengdqcp}
\bibinfo{author}{\bibfnamefont{J.}~\bibnamefont{Zhao}},
  \bibinfo{author}{\bibfnamefont{Y.-C.} \bibnamefont{Wang}},
  \bibinfo{author}{\bibfnamefont{Z.}~\bibnamefont{Yan}},
  \bibinfo{author}{\bibfnamefont{M.}~\bibnamefont{Cheng}}, \bibnamefont{and}
  \bibinfo{author}{\bibfnamefont{Z.~Y.} \bibnamefont{Meng}},
  \bibinfo{journal}{Phys. Rev. Lett.} \textbf{\bibinfo{volume}{128}},
  \bibinfo{pages}{010601} (\bibinfo{year}{2022}),
  \urlprefix\url{https://link.aps.org/doi/10.1103/PhysRevLett.128.010601}.

\bibitem[{\citenamefont{Kos et~al.}(2015)\citenamefont{Kos, Poland,
  Simmons-Duffin, and Vichi}}]{bootstrap}
\bibinfo{author}{\bibfnamefont{F.}~\bibnamefont{Kos}},
  \bibinfo{author}{\bibfnamefont{D.}~\bibnamefont{Poland}},
  \bibinfo{author}{\bibfnamefont{D.}~\bibnamefont{Simmons-Duffin}},
  \bibnamefont{and} \bibinfo{author}{\bibfnamefont{A.}~\bibnamefont{Vichi}},
  \bibinfo{journal}{Journal of High energy Physics}
  \textbf{\bibinfo{volume}{1511}}, \bibinfo{pages}{106} (\bibinfo{year}{2015}).

\bibitem[{\citenamefont{Nakayama and Ohtsuki}(2016)}]{bootstrap1}
\bibinfo{author}{\bibfnamefont{Y.}~\bibnamefont{Nakayama}} \bibnamefont{and}
  \bibinfo{author}{\bibfnamefont{T.}~\bibnamefont{Ohtsuki}},
  \bibinfo{journal}{Phys. Rev. Lett.} \textbf{\bibinfo{volume}{117}},
  \bibinfo{pages}{131601} (\bibinfo{year}{2016}),
  \urlprefix\url{https://link.aps.org/doi/10.1103/PhysRevLett.117.131601}.

\bibitem[{\citenamefont{Sandvik}(2010{\natexlab{b}})}]{sandvik1}
\bibinfo{author}{\bibfnamefont{A.~W.} \bibnamefont{Sandvik}},
  \bibinfo{journal}{Phys. Rev. Lett.} \textbf{\bibinfo{volume}{104}},
  \bibinfo{pages}{137204} (\bibinfo{year}{2010}{\natexlab{b}}),
  \urlprefix\url{https://link.aps.org/doi/10.1103/PhysRevLett.104.137204}.

\bibitem[{\citenamefont{{Yang} et~al.}(2020)\citenamefont{{Yang}, {Yao}, and
  {Sandvik}}}]{sandvik2}
\bibinfo{author}{\bibfnamefont{S.}~\bibnamefont{{Yang}}},
  \bibinfo{author}{\bibfnamefont{D.-X.} \bibnamefont{{Yao}}}, \bibnamefont{and}
  \bibinfo{author}{\bibfnamefont{A.~W.} \bibnamefont{{Sandvik}}},
  \bibinfo{journal}{arXiv e-prints} \bibinfo{eid}{arXiv:2001.02821}
  (\bibinfo{year}{2020}), \eprint{2001.02821}.

\bibitem[{\citenamefont{{Grover} and {Vishwanath}}(2012)}]{groveredge}
\bibinfo{author}{\bibfnamefont{T.}~\bibnamefont{{Grover}}} \bibnamefont{and}
  \bibinfo{author}{\bibfnamefont{A.}~\bibnamefont{{Vishwanath}}},
  \bibinfo{journal}{arXiv e-prints} \bibinfo{eid}{arXiv:1206.1332}
  (\bibinfo{year}{2012}), \eprint{1206.1332}.

\bibitem[{\citenamefont{Zhang and Wang}(2017)}]{zhang1}
\bibinfo{author}{\bibfnamefont{L.}~\bibnamefont{Zhang}} \bibnamefont{and}
  \bibinfo{author}{\bibfnamefont{F.}~\bibnamefont{Wang}},
  \bibinfo{journal}{Phys. Rev. Lett.} \textbf{\bibinfo{volume}{118}},
  \bibinfo{pages}{087201} (\bibinfo{year}{2017}),
  \urlprefix\url{https://link.aps.org/doi/10.1103/PhysRevLett.118.087201}.

\bibitem[{\citenamefont{Ding et~al.}(2018)\citenamefont{Ding, Zhang, and
  Guo}}]{zhang2}
\bibinfo{author}{\bibfnamefont{C.}~\bibnamefont{Ding}},
  \bibinfo{author}{\bibfnamefont{L.}~\bibnamefont{Zhang}}, \bibnamefont{and}
  \bibinfo{author}{\bibfnamefont{W.}~\bibnamefont{Guo}},
  \bibinfo{journal}{Phys. Rev. Lett.} \textbf{\bibinfo{volume}{120}},
  \bibinfo{pages}{235701} (\bibinfo{year}{2018}),
  \urlprefix\url{https://link.aps.org/doi/10.1103/PhysRevLett.120.235701}.

\bibitem[{\citenamefont{Weber et~al.}(2018)\citenamefont{Weber, Parisen~Toldin,
  and Wessel}}]{stefan1}
\bibinfo{author}{\bibfnamefont{L.}~\bibnamefont{Weber}},
  \bibinfo{author}{\bibfnamefont{F.}~\bibnamefont{Parisen~Toldin}},
  \bibnamefont{and} \bibinfo{author}{\bibfnamefont{S.}~\bibnamefont{Wessel}},
  \bibinfo{journal}{Phys. Rev. B} \textbf{\bibinfo{volume}{98}},
  \bibinfo{pages}{140403} (\bibinfo{year}{2018}),
  \urlprefix\url{https://link.aps.org/doi/10.1103/PhysRevB.98.140403}.

\bibitem[{\citenamefont{Weber and Wessel}(2019)}]{stefan2}
\bibinfo{author}{\bibfnamefont{L.}~\bibnamefont{Weber}} \bibnamefont{and}
  \bibinfo{author}{\bibfnamefont{S.}~\bibnamefont{Wessel}},
  \bibinfo{journal}{Phys. Rev. B} \textbf{\bibinfo{volume}{100}},
  \bibinfo{pages}{054437} (\bibinfo{year}{2019}),
  \urlprefix\url{https://link.aps.org/doi/10.1103/PhysRevB.100.054437}.

\bibitem[{\citenamefont{Xu et~al.}(2020)\citenamefont{Xu, Wu, Jian, and
  Xu}}]{edgexu1}
\bibinfo{author}{\bibfnamefont{Y.}~\bibnamefont{Xu}},
  \bibinfo{author}{\bibfnamefont{X.-C.} \bibnamefont{Wu}},
  \bibinfo{author}{\bibfnamefont{C.-M.} \bibnamefont{Jian}}, \bibnamefont{and}
  \bibinfo{author}{\bibfnamefont{C.}~\bibnamefont{Xu}}, \bibinfo{journal}{Phys.
  Rev. B} \textbf{\bibinfo{volume}{101}}, \bibinfo{pages}{184419}
  (\bibinfo{year}{2020}),
  \urlprefix\url{https://link.aps.org/doi/10.1103/PhysRevB.101.184419}.

\bibitem[{\citenamefont{Jian et~al.}(2021)\citenamefont{Jian, Xu, Wu, and
  Xu}}]{edgexu2}
\bibinfo{author}{\bibfnamefont{C.-M.} \bibnamefont{Jian}},
  \bibinfo{author}{\bibfnamefont{Y.}~\bibnamefont{Xu}},
  \bibinfo{author}{\bibfnamefont{X.-C.} \bibnamefont{Wu}}, \bibnamefont{and}
  \bibinfo{author}{\bibfnamefont{C.}~\bibnamefont{Xu}},
  \bibinfo{journal}{SciPost Phys.} \textbf{\bibinfo{volume}{10}},
  \bibinfo{pages}{33} (\bibinfo{year}{2021}),
  \urlprefix\url{https://scipost.org/10.21468/SciPostPhys.10.2.033}.

\bibitem[{\citenamefont{Metlitski}(2022)}]{maxboundary}
\bibinfo{author}{\bibfnamefont{M.~A.} \bibnamefont{Metlitski}},
  \bibinfo{journal}{SciPost Phys.} \textbf{\bibinfo{volume}{12}},
  \bibinfo{pages}{131} (\bibinfo{year}{2022}),
  \urlprefix\url{https://scipost.org/10.21468/SciPostPhys.12.4.131}.

\bibitem[{\citenamefont{Liu et~al.}(2021)\citenamefont{Liu, Shapourian,
  Vishwanath, and Metlitski}}]{shang1}
\bibinfo{author}{\bibfnamefont{S.}~\bibnamefont{Liu}},
  \bibinfo{author}{\bibfnamefont{H.}~\bibnamefont{Shapourian}},
  \bibinfo{author}{\bibfnamefont{A.}~\bibnamefont{Vishwanath}},
  \bibnamefont{and} \bibinfo{author}{\bibfnamefont{M.~A.}
  \bibnamefont{Metlitski}}, \bibinfo{journal}{Phys. Rev. B}
  \textbf{\bibinfo{volume}{104}}, \bibinfo{pages}{104201}
  (\bibinfo{year}{2021}),
  \urlprefix\url{https://link.aps.org/doi/10.1103/PhysRevB.104.104201}.

\bibitem[{\citenamefont{Parisen~Toldin}(2021)}]{toldin1}
\bibinfo{author}{\bibfnamefont{F.}~\bibnamefont{Parisen~Toldin}},
  \bibinfo{journal}{Phys. Rev. Lett.} \textbf{\bibinfo{volume}{126}},
  \bibinfo{pages}{135701} (\bibinfo{year}{2021}),
  \urlprefix\url{https://link.aps.org/doi/10.1103/PhysRevLett.126.135701}.

\bibitem[{\citenamefont{Parisen~Toldin and Metlitski}(2022)}]{toldin2}
\bibinfo{author}{\bibfnamefont{F.}~\bibnamefont{Parisen~Toldin}}
  \bibnamefont{and} \bibinfo{author}{\bibfnamefont{M.~A.}
  \bibnamefont{Metlitski}}, \bibinfo{journal}{Phys. Rev. Lett.}
  \textbf{\bibinfo{volume}{128}}, \bibinfo{pages}{215701}
  (\bibinfo{year}{2022}),
  \urlprefix\url{https://link.aps.org/doi/10.1103/PhysRevLett.128.215701}.

\bibitem[{\citenamefont{Ma et~al.}(2022)\citenamefont{Ma, Zou, and
  Wang}}]{maboundary}
\bibinfo{author}{\bibfnamefont{R.}~\bibnamefont{Ma}},
  \bibinfo{author}{\bibfnamefont{L.}~\bibnamefont{Zou}}, \bibnamefont{and}
  \bibinfo{author}{\bibfnamefont{C.}~\bibnamefont{Wang}},
  \bibinfo{journal}{SciPost Phys.} \textbf{\bibinfo{volume}{12}},
  \bibinfo{pages}{196} (\bibinfo{year}{2022}),
  \urlprefix\url{https://scipost.org/10.21468/SciPostPhys.12.6.196}.

\bibitem[{\citenamefont{Hertz}(1976)}]{hertz}
\bibinfo{author}{\bibfnamefont{J.~A.} \bibnamefont{Hertz}},
  \bibinfo{journal}{Phys. Rev. B} \textbf{\bibinfo{volume}{14}},
  \bibinfo{pages}{1165} (\bibinfo{year}{1976}),
  \urlprefix\url{https://link.aps.org/doi/10.1103/PhysRevB.14.1165}.

\bibitem[{\citenamefont{Millis}(1993)}]{millis}
\bibinfo{author}{\bibfnamefont{A.~J.} \bibnamefont{Millis}},
  \bibinfo{journal}{Phys. Rev. B} \textbf{\bibinfo{volume}{48}},
  \bibinfo{pages}{7183} (\bibinfo{year}{1993}),
  \urlprefix\url{https://link.aps.org/doi/10.1103/PhysRevB.48.7183}.

\bibitem[{\citenamefont{Senthil and Fisher}(2005)}]{senthilfisher}
\bibinfo{author}{\bibfnamefont{T.}~\bibnamefont{Senthil}} \bibnamefont{and}
  \bibinfo{author}{\bibfnamefont{M.~P.~A.} \bibnamefont{Fisher}},
  \bibinfo{journal}{Phys. Rev. B} \textbf{\bibinfo{volume}{74}},
  \bibinfo{pages}{064405} (\bibinfo{year}{2005}).

\bibitem[{\citenamefont{Kaul and Sachdev}(2008)}]{kaulsachdev}
\bibinfo{author}{\bibfnamefont{R.~K.} \bibnamefont{Kaul}} \bibnamefont{and}
  \bibinfo{author}{\bibfnamefont{S.}~\bibnamefont{Sachdev}},
  \bibinfo{journal}{Phys. Rev. B} \textbf{\bibinfo{volume}{77}},
  \bibinfo{pages}{155105} (\bibinfo{year}{2008}),
  \urlprefix\url{https://link.aps.org/doi/10.1103/PhysRevB.77.155105}.

\bibitem[{\citenamefont{Benvenuti and Khachatryan}(2019)}]{B2019}
\bibinfo{author}{\bibfnamefont{S.}~\bibnamefont{Benvenuti}} \bibnamefont{and}
  \bibinfo{author}{\bibfnamefont{H.}~\bibnamefont{Khachatryan}},
  \bibinfo{journal}{Journal of High Energy Physics}
  \textbf{\bibinfo{volume}{2019}}, \bibinfo{pages}{214} (\bibinfo{year}{2019}),
  ISSN \bibinfo{issn}{1029-8479},
  \urlprefix\url{https://doi.org/10.1007/JHEP05(2019)214}.

\bibitem[{\citenamefont{Wen and Wu}(1993)}]{wenwu}
\bibinfo{author}{\bibfnamefont{X.-G.} \bibnamefont{Wen}} \bibnamefont{and}
  \bibinfo{author}{\bibfnamefont{Y.-S.} \bibnamefont{Wu}},
  \bibinfo{journal}{Phys. Rev. Lett.} \textbf{\bibinfo{volume}{70}},
  \bibinfo{pages}{1501} (\bibinfo{year}{1993}),
  \urlprefix\url{https://link.aps.org/doi/10.1103/PhysRevLett.70.1501}.

\bibitem[{\citenamefont{Abanov and Chubukov}(2004)}]{hertzvalid}
\bibinfo{author}{\bibfnamefont{A.}~\bibnamefont{Abanov}} \bibnamefont{and}
  \bibinfo{author}{\bibfnamefont{A.}~\bibnamefont{Chubukov}},
  \bibinfo{journal}{Phys. Rev. Lett.} \textbf{\bibinfo{volume}{93}},
  \bibinfo{pages}{255702} (\bibinfo{year}{2004}),
  \urlprefix\url{https://link.aps.org/doi/10.1103/PhysRevLett.93.255702}.

\bibitem[{\citenamefont{Wang et~al.}(2017{\natexlab{b}})\citenamefont{Wang,
  Nahum, Metlitski, Xu, and Senthil}}]{SO5}
\bibinfo{author}{\bibfnamefont{C.}~\bibnamefont{Wang}},
  \bibinfo{author}{\bibfnamefont{A.}~\bibnamefont{Nahum}},
  \bibinfo{author}{\bibfnamefont{M.~A.} \bibnamefont{Metlitski}},
  \bibinfo{author}{\bibfnamefont{C.}~\bibnamefont{Xu}}, \bibnamefont{and}
  \bibinfo{author}{\bibfnamefont{T.}~\bibnamefont{Senthil}},
  \bibinfo{journal}{Phys. Rev. X} \textbf{\bibinfo{volume}{7}},
  \bibinfo{pages}{031051} (\bibinfo{year}{2017}{\natexlab{b}}),
  \urlprefix\url{https://link.aps.org/doi/10.1103/PhysRevX.7.031051}.

\bibitem[{\citenamefont{Metlitski and Sachdev}(2010)}]{metlitskisdw}
\bibinfo{author}{\bibfnamefont{M.~A.} \bibnamefont{Metlitski}}
  \bibnamefont{and} \bibinfo{author}{\bibfnamefont{S.}~\bibnamefont{Sachdev}},
  \bibinfo{journal}{Phys. Rev. B} \textbf{\bibinfo{volume}{82}},
  \bibinfo{pages}{075128} (\bibinfo{year}{2010}),
  \urlprefix\url{https://link.aps.org/doi/10.1103/PhysRevB.82.075128}.

\bibitem[{\citenamefont{Schlief et~al.}(2017)\citenamefont{Schlief, Lunts, and
  Lee}}]{leesdw}
\bibinfo{author}{\bibfnamefont{A.}~\bibnamefont{Schlief}},
  \bibinfo{author}{\bibfnamefont{P.}~\bibnamefont{Lunts}}, \bibnamefont{and}
  \bibinfo{author}{\bibfnamefont{S.-S.} \bibnamefont{Lee}},
  \bibinfo{journal}{Phys. Rev. X} \textbf{\bibinfo{volume}{7}},
  \bibinfo{pages}{021010} (\bibinfo{year}{2017}),
  \urlprefix\url{https://link.aps.org/doi/10.1103/PhysRevX.7.021010}.

\bibitem[{\citenamefont{Xu and Ludwig}(2013)}]{xuludwig}
\bibinfo{author}{\bibfnamefont{C.}~\bibnamefont{Xu}} \bibnamefont{and}
  \bibinfo{author}{\bibfnamefont{A.~W.~W.} \bibnamefont{Ludwig}},
  \bibinfo{journal}{Phys. Rev. Lett.} \textbf{\bibinfo{volume}{110}},
  \bibinfo{pages}{200405} (\bibinfo{year}{2013}),
  \urlprefix\url{https://link.aps.org/doi/10.1103/PhysRevLett.110.200405}.

\bibitem[{\citenamefont{Bi et~al.}(2016)\citenamefont{Bi, Rasmussen, BenTov,
  and Xu}}]{xunlsm}
\bibinfo{author}{\bibfnamefont{Z.}~\bibnamefont{Bi}},
  \bibinfo{author}{\bibfnamefont{A.}~\bibnamefont{Rasmussen}},
  \bibinfo{author}{\bibfnamefont{Y.}~\bibnamefont{BenTov}}, \bibnamefont{and}
  \bibinfo{author}{\bibfnamefont{C.}~\bibnamefont{Xu}},
  \emph{\bibinfo{title}{Stable interacting (2 + 1)d conformal field theories at
  the boundary of a class of (3 + 1)d symmetry protected topological phases}}
  (\bibinfo{year}{2016}), \urlprefix\url{https://arxiv.org/abs/1605.05336}.

\bibitem[{\citenamefont{Ma and Wang}(2020)}]{wangnlsm}
\bibinfo{author}{\bibfnamefont{R.}~\bibnamefont{Ma}} \bibnamefont{and}
  \bibinfo{author}{\bibfnamefont{C.}~\bibnamefont{Wang}},
  \bibinfo{journal}{Phys. Rev. B} \textbf{\bibinfo{volume}{102}},
  \bibinfo{pages}{020407} (\bibinfo{year}{2020}),
  \urlprefix\url{https://link.aps.org/doi/10.1103/PhysRevB.102.020407}.

\bibitem[{\citenamefont{Nahum}(2020)}]{nahumlsm}
\bibinfo{author}{\bibfnamefont{A.}~\bibnamefont{Nahum}},
  \bibinfo{journal}{Phys. Rev. B} \textbf{\bibinfo{volume}{102}},
  \bibinfo{pages}{201116} (\bibinfo{year}{2020}),
  \urlprefix\url{https://link.aps.org/doi/10.1103/PhysRevB.102.201116}.

\bibitem[{\citenamefont{Wang et~al.}(2021)\citenamefont{Wang, Zaletel, Mong,
  and Assaad}}]{assaadnlsm}
\bibinfo{author}{\bibfnamefont{Z.}~\bibnamefont{Wang}},
  \bibinfo{author}{\bibfnamefont{M.~P.} \bibnamefont{Zaletel}},
  \bibinfo{author}{\bibfnamefont{R.~S.~K.} \bibnamefont{Mong}},
  \bibnamefont{and} \bibinfo{author}{\bibfnamefont{F.~F.}
  \bibnamefont{Assaad}}, \bibinfo{journal}{Phys. Rev. Lett.}
  \textbf{\bibinfo{volume}{126}}, \bibinfo{pages}{045701}
  (\bibinfo{year}{2021}),
  \urlprefix\url{https://link.aps.org/doi/10.1103/PhysRevLett.126.045701}.

\bibitem[{\citenamefont{Zou et~al.}(2021)\citenamefont{Zou, He, and
  Wang}}]{stiefel}
\bibinfo{author}{\bibfnamefont{L.}~\bibnamefont{Zou}},
  \bibinfo{author}{\bibfnamefont{Y.-C.} \bibnamefont{He}}, \bibnamefont{and}
  \bibinfo{author}{\bibfnamefont{C.}~\bibnamefont{Wang}},
  \bibinfo{journal}{Phys. Rev. X} \textbf{\bibinfo{volume}{11}},
  \bibinfo{pages}{031043} (\bibinfo{year}{2021}),
  \urlprefix\url{https://link.aps.org/doi/10.1103/PhysRevX.11.031043}.

\bibitem[{\citenamefont{Xu et~al.}(2021)\citenamefont{Xu, Wu, Ye, Luo, Jian,
  and Xu}}]{fractionalMIT}
\bibinfo{author}{\bibfnamefont{Y.}~\bibnamefont{Xu}},
  \bibinfo{author}{\bibfnamefont{X.-C.} \bibnamefont{Wu}},
  \bibinfo{author}{\bibfnamefont{M.}~\bibnamefont{Ye}},
  \bibinfo{author}{\bibfnamefont{Z.-X.} \bibnamefont{Luo}},
  \bibinfo{author}{\bibfnamefont{C.-M.} \bibnamefont{Jian}}, \bibnamefont{and}
  \bibinfo{author}{\bibfnamefont{C.}~\bibnamefont{Xu}},
  \emph{\bibinfo{title}{Metal-insulator transition with charge
  fractionalization}} (\bibinfo{year}{2021}),
  \urlprefix\url{https://arxiv.org/abs/2106.14910}.

\bibitem[{\citenamefont{Musser et~al.}(2021)\citenamefont{Musser, Senthil, and
  Chowdhury}}]{wignerMIT}
\bibinfo{author}{\bibfnamefont{S.}~\bibnamefont{Musser}},
  \bibinfo{author}{\bibfnamefont{T.}~\bibnamefont{Senthil}}, \bibnamefont{and}
  \bibinfo{author}{\bibfnamefont{D.}~\bibnamefont{Chowdhury}},
  \emph{\bibinfo{title}{Theory of a continuous bandwidth-tuned wigner-mott
  transition}} (\bibinfo{year}{2021}),
  \urlprefix\url{https://arxiv.org/abs/2111.09894}.

\bibitem[{\citenamefont{Senthil}(2008)}]{senthilMIT}
\bibinfo{author}{\bibfnamefont{T.}~\bibnamefont{Senthil}},
  \bibinfo{journal}{Phys. Rev. B} \textbf{\bibinfo{volume}{78}},
  \bibinfo{pages}{045109} (\bibinfo{year}{2008}),
  \urlprefix\url{https://link.aps.org/doi/10.1103/PhysRevB.78.045109}.

\end{thebibliography}
	
\end{document}